# Synthesis and thermal expansion of chalcogenide MAX phase Hf$_2$SeC

*Xudong Wang* [a,b,#], *Ke Chen* [a,c,#,*], *Erxiao Wu* [a], *Yiming Zhang* [a,c], *Haoming Ding* [a], *Nianxiang Qiu* [a,c], *Yujie Song* [a,c], *Shiyu Du* [a,c], *Zhifang Chai* [a,c], *Qing Huang* [a,c,*]

[a] Engineering Laboratory of Advanced Energy Materials, Ningbo Institute of Materials Technology and Engineering, Chinese Academy of Sciences, Ningbo 315201, China.

[b] School of Materials Science and Engineering, Shanghai University, Shanghai 200444, China

[c] Qianwan Institute of CNiTECH, Ningbo 315336, China.

[#] These authors contributed equally to this work.

**\*** Corresponding authors.

E-mail addresses: chenke@nimte.ac.cn (Ke Chen), huangqing@nimte.ac.cn (Qing Huang).

**Abstract**

Thermal expansion of MAX phases along different directions tended to be different because of the anisotropy of hexagonal crystals. Herein, a new Hf$_2$SeC phase was synthesized and confirmed to be relatively isotropic, whose coefficients of thermal expansion (CTEs) were determined to be 9.73 μK$^{-1}$ and 10.18 μK$^{-1}$ along *a* and *c* directions. The strong M-S bond endowed Hf$_2$SC and Zr$_2$SC lower CTEs than Hf$_2$SeC and Zr$_2$SeC. A good relationship between the thermal expansion anisotropy and the ratio of elastic stiffness constant $c_{11}$ and $c_{33}$ was established. This straightforward approximation could be used to roughly predict the thermal expansion anisotropy of MAX phases.

**Key wards:** MAX phase, Hf$_2$SeC, thermal expansion, anisotropy

# 1. Introduction

MAX phases emerged in 1960s, are a large family of layered ternary compounds [1]. These early transition-metal carbides and nitrides have a general chemical formula $M_{n+1}AX_n$; wherein M is an early transition metal, A mainly comes from A-group element, X is carbon and/or nitrogen, n = 1 – 3 [1, 2]. Due to the combining properties of both metals and ceramics, MAX phases have good high-temperature stability, good damage tolerant, high thermal conductivity, as well as excellent resistance to oxidation and corrosion [1-4]. Most of the physical and chemical properties of MAX phases are anisotropic because of the layered, hexagonal crystal structure, whose edge-sharing covalent $M_6X$ octahedrons are interleaved by close-packed metallic A layers [1, 2]. However, anisotropic crystal sometimes is harmful to the applications of MAX phases, especially in extreme environments including high temperature and intense irradiation fields. For instance, the grain boundary cracking induced by energetic heavy ion (5.8 MeV Ni) could be manifested by the anisotropic swelling [5]. Hence, the relatively isotropic $Ti_3SiC_2$ with similar elastic modulus along *a* and *c* directions exhibited better irradiation damage tolerant than aluminum-based MAX phases ($Ti_3AlC_2$ and $Ti_2AlC$) [5, 6].

$Ti_2SC$ is compact in crystal structure, having a much smaller *c/a* value and stronger M-A bond compared with the other 211-MAX phases [7, 8]. Hence, the thermal expansion anisotropy of $Ti_2SC$ is relatively low, whose coefficients of thermal expansions (CTE) along *a* and *c* directions were 8.6 $\mu K^{-1}$ and 8.7 $\mu K^{-1}$ [6, 9]. The high bond strength of M-A can also be found in the newly synthesized chalcogenide MAX phase, $Zr_2SeC$. The calculation of density of states confirms the strong interaction between outer electrons of Zr and that of Se [10, 11]. Herein, $Hf_2SeC$ was successfully prepared to further explore the thermal expansion anisotropy of this series of new Se-MAX phases. The expanded crystal parameters of $Hf_2SeC$ powders were measured and calculated by high-temperature X-ray diffraction and subsequent Rietveld refinement from ambient temperature to 600 °C using $Hf_2SC$, $Zr_2SeC$, and $Zr_2SC$ as comparative analyses. The elastic stiffness constants were calculated to plot the influence of A-site elements on the thermal expansion anisotropy of these chalcogenide MAX phases.

## 2. Experimental

### *2.1 Synthesis details*

To begin with, considering the low melting and boiling points of selenium, the hafnium diselenide was synthesized using commercially hafnium (99.9% purity, 400 nm, Beijing Xingrongyuan Technology Co., Ltd. China) and selenium (99.99% purity, 200 mesh, Aladdin) powders in a sealed quartz tube at 950°C for 24 hours [12]. $Hf_2SeC$ was synthesized using the mixture of hafnium, self-made hafnium diselenide and carbon (99.9% purity, 400 nm, Macklin, China) powders with a ratio of 3:1.05:1.9. The ground powder mixture was pre-pressed in a graphite die with Φ20 mm. The reaction process was performed in a Pulse-Electric-Current-Aided sintering device (HPD 25/3, FCT Group, Germany) at the temperature of 1600 °C for 10 min with a maximum uniaxial pressure of 48 MPa under Ar atmosphere. In whole reaction process, the initial heating rate was 50 °C/min in the temperature range of 450-1100°C and reduced to 25 °C/min after 1100°C, and the cooling rate was 50°C/min in the temperature range of 1600-450°C.

In addition, $Hf_2SC$, $Zr_2SeC$ and $Zr_2SC$, were also prepared for comparative analyses. The precursors $HfS_2$, $ZrSe_2$ and $ZrS_2$ were synthesized at 950°C, 900°C and 950°C in the same method [12, 13]. It was worth reminding that the raw materials sulfur (99.99% purity, 300mesh, General Research Institute for Nonferrous Metals, China) and zirconium (99.5% purity, 400mesh, Targets Research Center of General Research Institute for Nonferrous Metals, China) were commercially available powders. The $Hf_2SC$, $Zr_2SeC$ and $Zr_2SC$ phases were *in-situ* synthesized using self-made transition-metal dichalcogenide, commercially metal and carbon powders with a stoichiometric ratio of 1.05:2.9:1.9, 1.05:3:1.9 [10] and 1:3:2 [14]. The reaction process was the same as the method for $Hf_2SeC$ except the sintering temperature of 1500°C. Finally, the as-obtained bulks were ground with 1 μm-diamond grinding paste for subsequent characterization.

### *2.2 Materials characterization*

The as-synthesized powders were examined and analyzed using X-ray diffractometer

(XRD; Bruker AXS D8 Advance, Germany) with Cu K$_α$ ray under an operating power of 1600 W (40 mA and 40 kV) at a step scan of 0.02 °/2$θ$ and a step time of 0.2s. The lattice parameters and phase composition of the synthesized phases were determined by the Rietveld method using the software of *TOPAS-Academic v6*. The fracture micromorphology and chemical composition of the specimens were studied by the scanning electron microscope (SEM; Hitachi Regulus 8230, Japan) equipped with energy dispersive spectroscopy (EDS). Furthermore, thin foils were prepared by focused ion beam (FIB; Thermo scientific Helios-G4-CX, USA) technique and were observed furtherly by a transmission electron microscope (TEM; Spectra 300, USA).

Coefficients of linear expansion (CTEs) in the *a* and *c* directions of the as-synthesized MAX phases powders were studied by the high-temperature XRD diffractometer (XRD; Bruker D8 Advance, Germany) with a furnace in vacuum ($10^{-2}$ mbar). The XRD data were taken from 5°–70° in steps of 0.02° withhold times of 0.2 s/step from room temperature to 873K. The heating rate was 60 °C/min with a 10 min holding time prior to each scan. Rietveld analysis was performed on the high-temperature XRD patterns using the *TOPAS-Academic v6* to calculate the expanded lattice parameters.

*2.3 Calculation Details*

The calculations were performed by the *Vienna Ab initio Computing Software Package* [15, 16], and the projector-augmented wave method [17] with an energy cutoff of 500 eV. The Perdew–Burke–Ernzerhof (PBE) version of the generalized gradient approximation (GGA) [18] was used for the exchange and correlation functionals. In order to keep the forces on each atom of the crystal structure model less than $1.0 × 10^{-4}$ eV/Å and the total energy converged at $1.0 × 10^{-8}$ eV/cell, the crystal structure was relaxed according to the conjugate gradient scheme [19].

The Brillouin Zones (BZs) of MAX phases configurations were described with a Γ-centered *k*-point mesh of 12 × 12 × 4. The density of states was plotted by the Gaussian smearing with a broadening of 0.01 eV. The bulk states were optimized through relaxing all the volume, lattice parameters, and atomic positions. For crystals with hexagonal symmetry, elastic stiffness constants are calculated based on the stress–strain relationship [20].

## 3. Results and discussion

The as-synthesized bulks were characterized using X-ray diffraction from 5 ° to 70 ° (**Figure 1(a)**). Rietveld fitting results (reliability factors $R_{wp}$ = 8.29 %) showed that the crystal structure of predominant phase belonged to hexagonal structure (P6$_3$/mmc, No. 194), which was the characteristic structure of MAX phase (**the inset in Figure 1(a)**) [1]. The determined composition was 95.2 wt. % Hf$_2$SeC with small amount of residual HfO$_2$. The respective coordinates of component elements in this chalcogenide MAX phase were Hf (1/3, 2/3, 0.0953), Se (1/3, 2/3, 3/4), and C (0, 0, 0), and fitting lattice parameters were $a$ = 0.3422 nm and $c$ = 1.2391 nm, respectively. This lattice parameter was in good agreement with the analysis of TEM. Wherein, the value of $a$ and $c$ were identified to be 0.3387 nm and 1.2327 nm according to the selected area electron diffraction (SAED) (**the inset in Figure 1(c)**). The first-principles calculations were also performed to verify the formation of Hf$_2$SeC MAX phase. The calculated lattice parameters were $a$ = 0.3436 nm and $c$ = 1.2452, and $z$ position in coordinate of Hf atom was 0.0943. The fracture morphology and composition of the as-synthesized products were characterized by SEM (**Figure 1(b)**). The hexagonal grains with size of about 1 μm were observed indicating the intergranular fracture of the as-synthesized Hf$_2$SeC bulk. EDS results showed that the composition of the main particles contained Hf, Se and C. The ratio of Hf to Se was close to 2: 1 (**Table S1**, **Figure S1**), indicating the formation of Hf$_2$SeC. The crystal structure and atom arrangements within the Hf$_2$SeC along the $[11\bar{2}0]$ zone axis was visually observed via advanced atom-resolved high-angle annular dark-field (HAADF) technique (**Figure 1(c)**). The HADDF image is related to atomic number contrast, *i.e.*, the larger the atomic number is, the brighter the atom will be [21]. It was easily to distinguish that two lines of bigger and brighter dots (Hf, also marked in crimson) interleaved with one lines of smaller and dimmer dots (Se, also marked in cyan) (**Figure 1(d-f)**). This mirrored zig-zag arrangement was the characteristic of 211 MAX phase [1, 3].

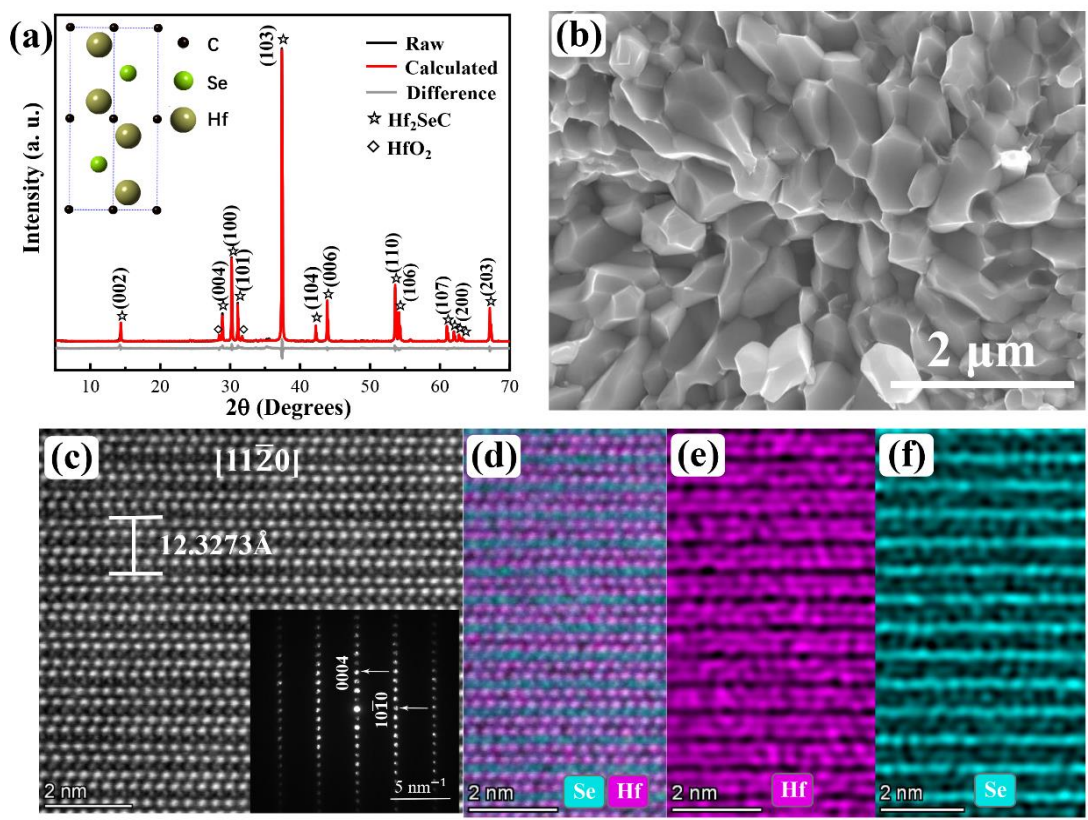

**Figure 1. (a)** XRD patterns of as-synthesized Hf$_2$SeC. **(b)** SEM image of the Hf$_2$SeC. **(c)** High-resolution image of the Hf$_2$SeC showing atomic positions along [11$\bar{2}$0] direction, the inset was corresponding selected area electron diffraction. **(d-f)** Atom-resolved EDS mapping images.

Thermal expansion is one of the important thermal properties for the high-temperature applications, which is induced by the anharmonic vibration of atoms in temperature field. Herein, the expanded lattice parameters of as-synthesized Hf$_2$SeC powders were measured by high-temperature X-ray diffraction from ambient temperature to 873 K (**Figure 2(a-c)**). It was clearly that the Hf$_2$SeC peaks noticeably shifted to the left with increasing temperature, which indicated the expansion of lattice structure. Among them, the evolutions of (103) peaks and (002) peaks were the typical to show the thermal expansion of MAX phases along *a* and *c* directions. The expanded lattice parameters were then determined by Rietveld refinements using *TOPAS-Academic v6*. The evolutions of refined lattice parameters were linearly fitted using Zr$_2$SeC, Hf$_2$SC, and Zr$_2$SC as comparative analyses (**Figure 2(d-e), Figure S2** and

**Table S2**). The coefficients of thermal expansion (CTEs) along *a* and *c* directions ($\alpha_a$ and $\alpha_c$) were calculated as follows [22]:

$$\alpha_a = \frac{d(a(T))}{a_0 dT} \text{ and } \alpha_c = \frac{d(c(T))}{c_0 dT} \quad (1)$$

Where, $\frac{d(a(T))}{dT}$ and $\frac{d(c(T))}{dT}$ are the slopes of fitted linear functions of *a(T)* and *c(T)*. $a_0$ and $c_0$ are the lattice parameters at ambient temperature. The mean CTE ($\alpha_{av}$) was then calculated as follows [22]:

$$\alpha_{av} = \frac{2\alpha_a + \alpha_c}{3} \quad (2)$$

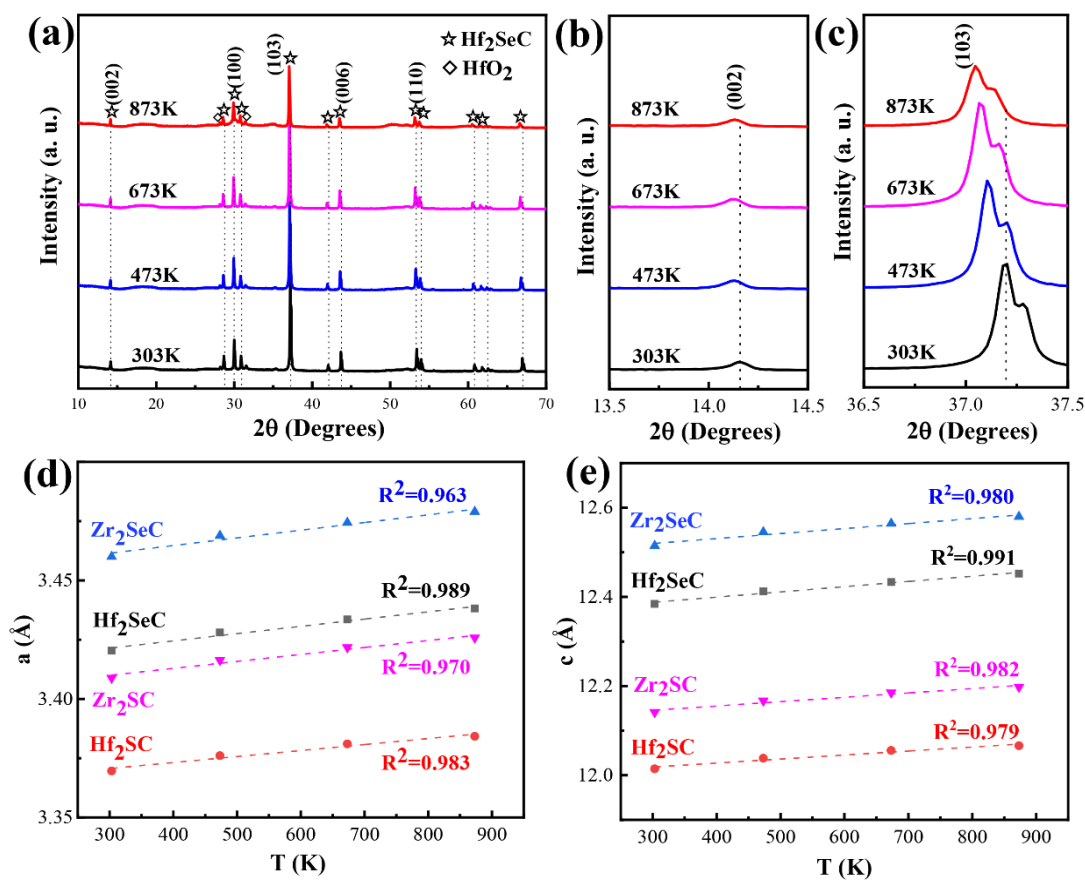

**Figure 2. (a-c)** High-temperature XRD patterns of Hf$_2$SeC. **(d-e)** Temperature dependence of the lattice parameters *a* and *c* for Hf$_2$SeC, Hf$_2$SC, Zr$_2$SeC and Zr$_2$SC.

Herein, $\alpha_a$, $\alpha_c$, and $\alpha_{av}$ of Hf$_2$SeC were given to be 9.73 μK$^{-1}$, 10.18 μK$^{-1}$, and 9.88 μK$^{-1}$, respectively. And CTEs of Zr$_2$SeC, Hf$_2$SC, and Zr$_2$SC were shown in **Table 1**. Interestingly, the CTEs of Se-MAX phases were larger than those of S-MAX phases, which indicated that the A-group elements played an important role in the thermal expansion of MAX phase. This discrepancy could be tentatively explained by the regulation of M–A bond length in these nanolaminated carbides. Compared with sulfur, selenium is larger in atomic size and has a smaller electronegativity. The occupation of Se in A site lengthened the M-A bond length of M-S, which was verified by first-principles calculation. Wherein, the bond length for Hf-Se and Zr-Se were 2.774 Å and 2.800 Å, while the bond length for Hf-S, and Zr-S were 2.662 Å and 2.691 Å, respectively (**Table S3**). This weakened bonding energy of Zr-Se than that of Zr-S was also confirmed by the analysis of density of states in the previous works [10, 11]. The soft crystal structure benefited the amplitude of vibration as reflected a high thermal expansion in $Hf_2SeC$ and $Zr_2SeC$.

**Table 1.** Coefficients of thermal expansion and thermal expansion anisotropy for $Hf_2SeC$, $Hf_2SC$, $Zr_2SeC$ and $Zr_2SC$.

| Compound | $\alpha_a$ (μK$^{-1}$) | $\alpha_c$ (μK$^{-1}$) | $\alpha_c/\alpha_a$ | $\alpha_{av}$ (μK$^{-1}$) |
|---|---|---|---|---|
| $Hf_2SeC$[a] | 9.73 | 10.18 | 1.05 | 9.88 |
| $Hf_2SC$[a] | 7.51 | 7.49 | 1.00 | 7.50 |
| $Zr_2SeC$[a] | 9.36 | 8.94 | 0.95 | 9.22 |
| $Zr_2SC$[a] | 8.61 | 8.04 | 0.93 | 8.42 |
| $Ti_2SC$[b] | 8.61 | 8.72 | 1.01 | 8.65 |

[a] Present work

[b] Reference [6]

**Table 2.** The Elastic stiffness constants and subsequently calculated Grüneisen parameters for $Hf_2SeC$, $Hf_2SC$, $Zr_2SeC$ and $Zr_2SC$.

| Compound | $c_{11}$ (GPa) | $c_{12}$ (GPa) | $c_{13}$ (GPa) | $c_{33}$ (GPa) | $c_{44}$ (GPa) | $\gamma_c/\gamma_a$ |
|---|---|---|---|---|---|---|
| $Hf_2SeC$[a] | 308 | 93 | 105 | 314 | 135 | 1.05 |
| $Hf_2SC$[a] | 330 | 100 | 118 | 345 | 150 | 1.06 |
| $Zr_2SeC$[a] | 275 | 82 | 92 | 289 | 125 | 1.03 |
| $Zr_2SC$[a] | 295 | 91 | 100 | 318 | 135 | 1.04 |
| $Ti_2SC$[b] | 339 | 90 | 100 | 354 | 162 | 1.05 |

[a] Present work

[b] Reference [6]

It was noteworthy that the $\alpha_c$ and $\alpha_a$ of $Hf_2SeC$, $Hf_2SC$, $Zr_2SeC$, and $Zr_2SC$ are almost the same (**Table 1**). The degree of anisotropy ratios for thermal expansion could be described as $\alpha_c/\alpha_a$. The values of $\alpha_c/\alpha_a$ for these four chalcogenide MAX phases were close to 1 (*i.e.*, $Hf_2SeC$, $Hf_2SC$, $Zr_2SeC$, and $Zr_2SC$ were 1.05, 1.00, 0.95, and 0.93 respectively), suggesting the isotropic expansion with the increasing temperature. This phenomenon is not too surprising since the bond stiffness of M-A and M-X associated with the anisotropy in thermal expansion are similar in the chalcogenide MAX phase [6, 11]. However, $Ti_2AlC$, as a common example of Al-MAX phase, had a very large difference between the bond stiffness of Ti-Al and Ti-C [24], whose value of $\alpha_c/\alpha_a$ was 1.41 [8]. In order to further discuss the influence of chalcogenide A-site elements on the anisotropy in thermal expansion, these four values of $\alpha_c/\alpha_a$ were plotted in the conclusion of previous work (**Figure 3(a)**) [6]. It was interesting that the anisotropy of $\alpha_c/\alpha_a$ increased with the rising A-site group number of MAX phase, and fell down over the IVA group. The anisotropy points of as-synthesized chalcogenide MAX phases clustered near the line of $\alpha_c/\alpha_a = 1$. The phenomenon could be tentatively explained by the analysis of Grüneisen parameter, which quantified the anharmonic vibration of atoms [25]. In the hexagonal symmetry, the Grüneisen parameters along *a* and *c* directions ($\gamma_a$ and $\gamma_c$) were calculated as follows [6, 26]:

$$\gamma_a = \frac{V_m}{C_v}[(c_{11} + c_{12})\alpha_a + c_{13}\alpha_c] \tag{3}$$

$$\gamma_c = \frac{V_m}{C_v}(2c_{13}\alpha_a + c_{33}\alpha_c) \tag{4}$$

Where, $V_m$ is the molar volume. $C_v$ is the molar specific heat at constant volume, which is $3NR$ assumed as the application of Dulong–Petit law ($N$ is the number of atoms per unit cell and $R$ is the ideal gas constant). $c_{ij}$ is an elastic stiffness constant (five independent elastic stiffness constants $c_{11}$, $c_{12}$, $c_{13}$, $c_{33}$, and $c_{44}$ for hexagonal symmetry). This derivation implied that the thermal expansion was mainly associated with the elastic stiffness constants. The anisotropy of thermal expansion could also be described using the ratio of $\gamma_c$ to $\gamma_a$. The calculated values of $\gamma_a$ and $\gamma_c$ for Hf$_2$SeC, Hf$_2$SC, Zr$_2$SeC, and Zr$_2$SC were between 1 to 2, and the ratios of $\gamma_c$ to $\gamma_a$ were near 1, which indicated that the thermal expansion of these along $a$ and $c$ directions was almost the same in these chalcogenide MAX phases. For comparison, the largest thermal expansion anisotropy belonging to Nb$_2$AsC (**Figure 3(a)**). Scabarozi et al simplified the linear correlation between thermal expansion anisotropy with the elastic stiffness constant $c_{13}$ in 211-MAX phase. The anisotropy was concluded related to the bonding between the M−A elements [6]. Herein, the relationship between thermal expansion anisotropy and value of $c_{11}/c_{33}$ was plotted to further discuss the influence of A-site elements on the thermal expansion anisotropy of these chalcogenide MAX phases. The thermal expansion of MAX phase tended to isotropy when the value of $c_{11}/c_{33}$ was close to 1, since the $c_{11}$ and $c_{33}$ reflected the bonding strengths in the $a$ and $c$ directions of hexagonal crystals. This criterion was roughly effective for the most MAX phases concluded in **Figure 3(b)**. The as-synthesized chalcogenide MAX phases, Hf$_2$SeC, Hf$_2$SC, Zr$_2$SeC, and Zr$_2$SC, whose $c_{11}$ was similar to $c_{33}$, had relatively isotropic thermal expansion. While Al-MAX phases, such as Ti$_2$AlC and Ti$_3$AlC$_2$, were relatively anisotropic in thermal expansion due to the large difference of $c_{11}$ and $c_{33}$. The thermal expansion anisotropy of Nb$_2$SnC, Nb$_2$AsC, V$_2$GeC, and V$_2$AsC were abnormally large. The reason probably resulted from the large values of $c_{12}$ and $c_{13}$. Wherein, the differences between $c_{12}$ or $c_{13}$ and $c_{11}$ or $c_{33}$ were much larger than those of other MAX phases [6, 8, 27]. The above analyses means that the approximation is valid when the elastic stiffness constants normal to the expanded plain can be neglected comparing to those in the plain. This straightforward approximation could be used to roughly predict the thermal expansion anisotropy of MAX phases.

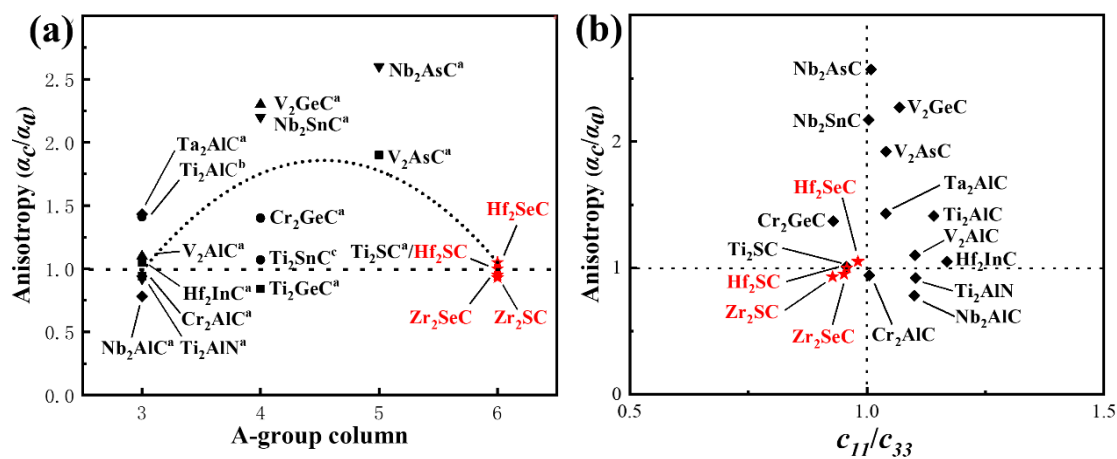

**Figure 3.** Thermal expansion properties of common MAX phases and the as-characterized four chalcogenide MAX phases were marked in red: **(a)** Anisotropy in thermal expansion vs A-group elements. **(b)** Anisotropy in thermal expansion vs elastic stiffness constant ratio of $c_{11}/c_{33}$.

[a] Reference [6]

[b] Reference [8]

[c] Reference [27]

## 4. Conclusions

In summary, a new chalcogenide MAX phase $Hf_2SeC$ was synthesized and characterized. The CTEs of $Hf_2SeC$ along $a$ and $c$ directions were determined to be 9.73 μK$^{-1}$, and 10.18 μK$^{-1}$, respectively. Due to the stronger M-A bond, S-MAX phases ($Hf_2SC$ and $Zr_2SC$) had lower CTEs than those of Se-MAX phases ($Hf_2SeC$ and $Zr_2SeC$). These four chalcogenide MAX phases were relatively isotropic in thermal expansion, which agreed with the approximation of correction between anisotropy and $c_{11}/c_{33}$.

**Declaration of Competing Interest**

No potential conflict of interest was reported by the authors.

**Acknowledgements**

This study was financially supported by the National Natural Science Foundation of China (Grant No. 51902319), International Partnership Program of Chinese Academy of Sciences (Grant No. 174433KYSB20190019), Leading Innovative and Entrepreneur Team Introduction Program of Zhejiang (Grant No. 2019R01003), Ningbo Top-talent Team Program and Ningbo "3315 plan" (Grant No. 2018A-03-A).

**Appendix A. Supplementary data**

Supplementary material related to this article can be found, in the online version, at https://doi.org/10.1016/j.jeurceramsoc.20xx.xx.xxx.

# Supplementary information

## Synthesis and thermal expansion of chalcogenide MAX phase Hf$_2$SeC

*Xudong Wang* [a,b,#], *Ke Chen* [a,c,#,*], *Erxiao Wu* [a], *Yiming Zhang* [a,c], *Haoming Ding* [a], *Nianxiang Qiu* [a,c], *Yujie Song* [a,c], *Shiyu Du* [a,c], *Zhifang Chai* [a,c], *Qing Huang* [a,c,*]

[a] Engineering Laboratory of Advanced Energy Materials, Ningbo Institute of Materials Technology and Engineering, Chinese Academy of Sciences, Ningbo 315201, China.

[b] School of Materials Science and Engineering, Shanghai University, Shanghai 200444, China

[c] Qianwan Institute of CNiTECH, Ningbo 315336, China.

[#] These authors contributed equally to this work.

* Corresponding authors.

E-mail addresses: chenke@nimte.ac.cn (Ke Chen), huangqing@nimte.ac.cn (Qing Huang).

**Figures and Tables**

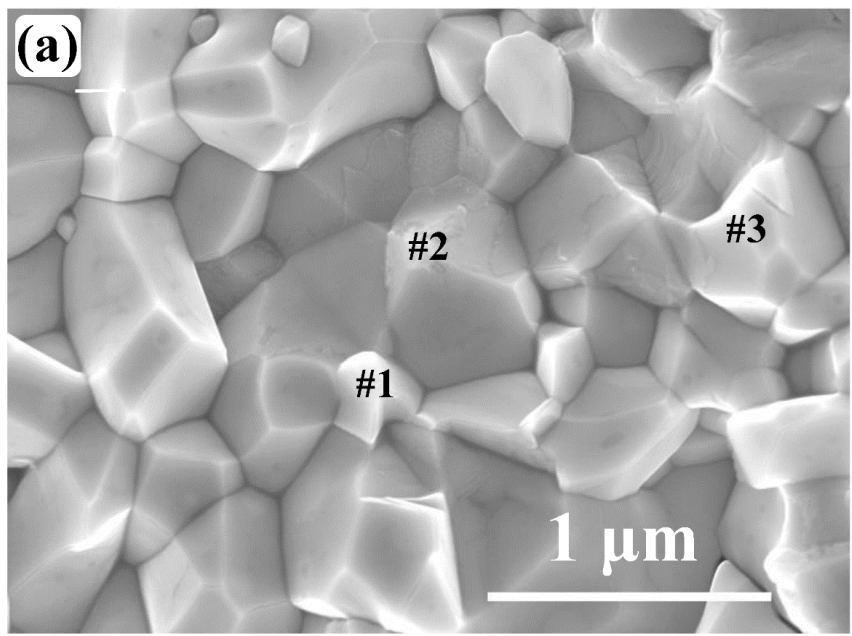

**Figure S1.** Fracture morphology of Hf$_2$SeC bulk combined with EDS point analysis.

Table S1. Elemental analysis of the selected area in **Figure S1**

| | Element | Hf | Se | C | O | Total |
|---|---|---|---|---|---|---|
| | #1 | 26.40 | 12.59 | 52.67 | 8.34 | 100.00 |
| at. % | #2 | 31.94 | 16.05 | 49.27 | 2.74 | 100.00 |
| | #3 | 28.30 | 13.78 | 56.28 | 1.63 | 99.99 |

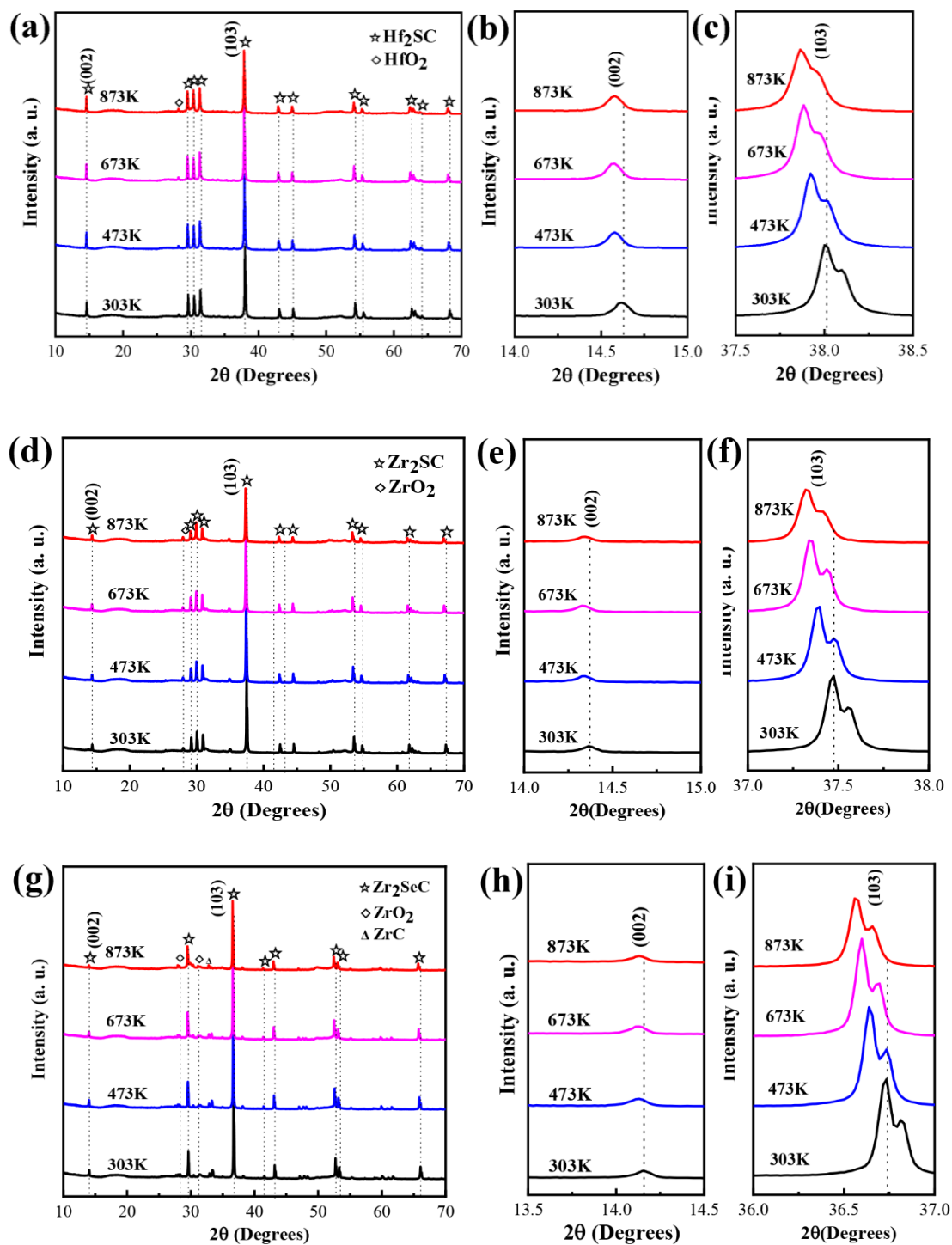

**Figure S2. (a-c)** High-temperature XRD patterns of Hf$_2$SC. **(d-f)** High-temperature XRD patterns of Zr$_2$SC. **(g-i)** High-temperature XRD patterns of Zr$_2$SeC.

**Table S2.** The expanded lattice parameters of four chalcogenide MAX phases using Rietveld refinements of high-temperature XRD patterns.

| Compound | $T$ (K) | $a$ (Å) | $c$ (Å) | $R_{wp}$ (%) |
|---|---|---|---|---|
| Hf$_2$SeC | 303 | 3.4204 | 12.3841 | 9.79 |
| | 473 | 3.4280 | 12.4127 | 10.72 |
| | 673 | 3.4335 | 12.4332 | 10.44 |
| | 873 | 3.4382 | 12.4523 | 11.01 |
| Hf$_2$SC | 303 | 3.3696 | 12.0143 | 9.48 |
| | 473 | 3.3761 | 12.0382 | 10.64 |
| | 673 | 3.3810 | 12.0556 | 10.56 |
| | 873 | 3.3842 | 12.0662 | 10.92 |
| Zr$_2$SeC | 303 | 3.4600 | 12.5145 | 9.83 |
| | 473 | 3.4689 | 12.5452 | 10.09 |
| | 673 | 3.4744 | 12.5644 | 10.92 |
| | 873 | 3.4790 | 12.5798 | 11.10 |
| Zr$_2$SC | 303 | 3.4088 | 12.1415 | 10.32 |
| | 473 | 3.4164 | 12.1671 | 10.76 |
| | 673 | 3.4218 | 12.1852 | 10.60 |
| | 873 | 3.4258 | 12.1980 | 10.64 |

**Table S3.** The bond length in crystal structures of $Hf_2SeC$, $Hf_2SC$, $Zr_2SeC$ and $Zr_2SC$.

| Compound | M-A (Å) | M-X (Å) |
|----------|---------|---------|
| $Hf_2SeC$ | 2.774 | 2.305 |
| $Hf_2SC$  | 2.662 | 2.291 |
| $Zr_2SeC$ | 2.800 | 2.342 |
| $Zr_2SC$  | 2.691 | 2.328 |